\documentclass[useAMS,usenatbib]{mn2e}
\usepackage{epsfig}
\usepackage{amsmath}
\usepackage{rotating}
\usepackage{longtable}
\usepackage{multirow}
\usepackage{amssymb}
\usepackage{graphicx}
\usepackage{graphics}
\usepackage{rotating}
\usepackage{lscape}
\usepackage{color}
\usepackage{todonotes}
\usepackage{float}

\newcommand{\be}{\begin{equation}}
\newcommand{\beq}{\begin{equation}}
\newcommand{\ba}{\begin{eqnarray}}
\newcommand{\ee}{\end{equation}}
\newcommand{\eeq}{\end{equation}}
\newcommand{\ea}{\end{eqnarray}}

\newcommand{\spitzer}{{\it Spitzer~}}

\newcommand{\NaID}{Na\,{\sc i\,D}\,}
\newcommand{\KI}{K\,{\sc i}\,}

\def\lsim{~\rlap{$<$}{\lower 1.0ex\hbox{$\sim$}}}

\def\gsim{~\rlap{$>$}{\lower 1.0ex\hbox{$\sim$}}}

\title[Spitzer observations of SN 2014J and SNe Ia]{Spitzer observations of SN 2014J and properties of mid-IR emission in Type Ia supernovae} 
\author[J. Johansson et al. ]
 	{J. Johansson$^1$\thanks{E-mail: joeljo@fysik.su.se},
	A. Goobar$^1$,
	M.~M. Kasliwal$^2$,
	G. Helou$^3$,
	F. Masci$^3$,
	S. Tinyanont$^4$,
	\newauthor J. Jencson$^2$,
	Y. Cao$^2$,
	O.~D. Fox$^5$,
	M. Kromer$^6$,
	R. Amanullah$^1$,
	D.~P.~K. Banerjee$^7$,
	\newauthor V. Joshi$^7$, 
	A. Jerkstrand$^8$,
	E. Kankare$^8$,
	T.~A. Prince$^2$
  \\$^1$ The Oskar Klein Centre, Department of Physics, Stockholm University, SE 106 91 Stockholm, Sweden 
  \\$^2$ Division of Physics, Mathematics, and Astronomy, California Institute of Technology, Pasadena, CA 91125, U.S.A. 
  \\$^3$ Infrared Processing and Analysis Center, California Institute of Technology, M/S 100-22, Pasadena, CA 91125, U.S.A.
  \\$^4$ IHarvey Mudd College, 301 Platt Boulevard, Claremont, CA 91711, U.S.A.
  \\$^5$ Department of Astronomy, University of California, Berkeley, CA 94720-3411, U.S.A.
  \\$^6$ The Oskar Klein Centre, Department of Astronomy, Stockholm University,  SE 106 91 Stockholm, Sweden 
  \\$^7$ Astronomy and Astrophysics Division, Physical Research Laboratory, Navrangpura, Ahmedabad, India 380009
  \\$^8$ Astrophysics Research Center, School of Mathematics and Physics, Queen's University Belfast, BT7 1NN, UK
 }
\date{Accepted 2016 December 20. Received 2016 December 1; in original form 2014 November 12}

\pagerange{\pageref{firstpage}--\pageref{lastpage}} \pubyear{2012}

\def\LaTeX{L\kern-.36em\raise.3ex\hbox{a}\kern-.15em
    T\kern-.1667em\lower.7ex\hbox{E}\kern-.125emX}

\begin{document}

\label{firstpage}

\maketitle

\begin{abstract}
  SN 2014J in M\,82 is the closest Type Ia supernova (SN~Ia) in decades.
  The proximity allows for detailed studies of supernova physics and provides
  insights into the circumstellar and interstellar environment.  In
  this work we analyze \spitzer mid-IR data of SN~2014J in the $3.6$
  and $4.5\,\mu$m wavelength range, together with several other nearby
  and well-studied SNe~Ia.  We compile the first composite mid-IR
  light-curve templates from our sample of SNe~Ia, spanning the range from before peak 
  brightness well into the nebular phase. Our observations indicate that SNe~Ia form a 
  very homogeneous class of objects at these wavelengths.  
  Using the low-reddening supernovae for comparison, we constrain possible thermal emission from
  circumstellar dust around the highly reddened SN~2014J. We also study SNe 2006X and 2007le, 
  where the presence of matter in the circumstellar environment has been suggested. No significant mid-IR excess 
  is detected, allowing us to place upper limits on the amount of pre-existing dust in the circumstellar environment. 
  For SN~2014J, $M_{\rm dust} \lsim 10^{-5} \, {\rm M}_{\odot}$ within $r_{\rm dust} \sim 10^{17}$\,cm, which is insufficient
  to account for the observed extinction.  Similar limits are obtained for SNe 2006X and 2007le.
\end{abstract}

\begin{keywords}
ISM: dust, extinction -- supernovae: general - circumstellar matter -- supernovae: individual: 2005df, 2006X, 2007af, 2007le, 2007sr, 2009ig, 2012cg, 2011fe, 2014J
\end{keywords}
%%%%%%%%%%%%%%%%%%%%%%%%%%%%%%%%%%%%%%%%%%%%%%%%%%%%%%%%%%%%%%%%%%%%%%%
%%%%%%%%%%%%%%%%%%%%%%%%%%%%%%%%%%%%%%%%%%%%%%%%%%%%%%%%%%%%%%%%%%%%%%%
\section{Introduction}
The use of Type Ia supernovae (SNe~Ia) as distance indicators remains
essential for the study of the expansion history of the Universe and
for exploring the nature of dark energy \citep[see e.g. review by ][]{goobar2011}.
However, a lack of understanding of the progenitor systems and the
requirement for empirically derived colour-brightness corrections
represent severe challenges for precision cosmology.  
SN~2014J in the starburst galaxy M\,82 is the closest SN~Ia in decades
and offers a unique opportunity to study 
progenitor and explosion models, as well as the circumstellar (CS) and interstellar (IS) medium
along the line-of-sight across an unprecedented wavelength range,
from gamma-rays to radio \citep{diehl2014,perez-torres2014}.

Our \spitzer observations of SN~2014J,
along with observations of other well-studied nearby SNe~Ia, allow us
to compile the first mid-IR light-curve templates, 
presenting us with a new wavelength range to confront theoretical models of SNe~Ia.
For example, the study of the mid-IR light-curve
around the near-IR secondary maximum, shown by \citet{kasen2006} to 
be a valuable diagnostic of the physical parameters governing SN~Ia explosions,
could be used to test the validity of the model predictions.

Furthermore, by searching for excess mid-IR emission towards SN~2014J,
we set constraints on the amounts of pre-existing
CS dust that could account for the non-standard reddening measured by
\citet[]{goobar2014a,amanullah2014,marion2014,foley2014,brown2014,ashall2014},
who otherwise find this SN to be a normal SN~Ia, albeit 
with slightly higher than average expansion velocities, and showing
signs of additional sources of luminosity in the first hours
after the explosion \citep{goobar2014b}. We also analyze other SNe~Ia with peculiar 
reddening observed with \spitzer, SNe~2006X and 2007le, where 
CS material has been reported \citep{patat2007,simon2009}.

The outline of this paper is as follows: 
In Section 2 we present the observations, followed by a description
in Section 3 of how the mid-IR light-curve templates are constructed. In Section 
4 we add this new wavelength range along with optical and near-IR spectra,
to present the spectral energy distribution (SED) in the full range
0.4 -- 5.0 $\mu$m. The implications for emission from heated CS dust
are discussed in Section 5, followed by concluding remarks in Section 6.
%%%%%%%%%%%%%%%%%%%%%%%%%%%%%%%%%%%%%%%%%%%%%%%%%%%%%%%%%%%%%%%%%%%%%%%
%%%%%%%%%%%%%%%%%%%%%%%%%%%%%%%%%%%%%%%%%%%%%%%%%%%%%%%%%%%%%%%%%%%%%%%
\section{Observations}
\subsection{SN 2014J}
SN~2014J was observed with \spitzer over 6 epochs between January 28
and March 3 and 4 epochs between May 31 and July 8 under the 
{\it SPitzer InfraRed Intensive Transients Survey} (SPIRITS) program (PI:
M. Kasliwal).  SPIRITS is an ongoing infrared survey that
systematically searches two hundred nearby galaxies for all types of
transients and variables within a volume of 20\,Mpc. 
Data is promptly processed with subtractions relative to
archival images in the Spitzer Heritage Archive. The SPIRITS team
undertakes a large concomitant ground-based survey in the optical and
near-IR to characterize the \spitzer findings. For additional details
about the survey and first discoveries, see \citet{Kasliwal2017}.

Aperture photometry was performed at the location of the SN on the
aligned \spitzer Post-Basic Calibrated Data for the SN and pre-SN images.
Throughout the paper we use the zero magnitude fluxes for the IRAC
Channels 1 and 2 (CH1 and CH2, with central wavelengths of 3.6 and 4.5
$\mu$m, respectively) of $F_{\nu,0}^{\rm CH1} = 280.9$\,Jy and
$F_{\nu,0}^{\rm CH2} = 179.7$\,Jy. All photometry is listed in Table~\ref{tab:photometry}.

We also present new optical and near-IR spectra (summarized in Table~\ref{tab:obslog}) 
and photometry (to be published in a future paper) of SN~2014J in Figs.~\ref{fig:lightcurves} and \ref{fig:sed}.

\subsection{Comparison Supernovae}\label{sec:sne}
In order to analyze our mid-IR data on SN~2014J we need to compare with other well-studied SNe~Ia.
For this study, we include SNe that have multi-epoch \spitzer data (see Table~\ref{tab:photometry}) and good optical/near-IR coverage.
{\bf SN~2011fe} was observed with \spitzer starting 145 days after $B-$band
maximum.  UV data \citep{mazzali2014} and optical to near-IR observations
\citep{pereira2013, matheson2012} together with high-resolution
spectroscopy \citep{patat2013}, show that SN~2011fe suffered little to
no extinction, $E(B-V) = 0.026 \pm 0.036$\,mag, making it useful as a
template of a pristine, normal SN~Ia.  We adopt a distance modulus to
SN~2011fe of $\mu = 28.93 \pm 0.16$\,mag ($D=6.1\pm0.45$\,Mpc) based on
near-IR light curves from \citet{matheson2012} which is in good agreement
with the Cepheid distance in \citet{2013ApJ...777...79M}.
{\bf SN~2012cg} was observed with \spitzer (PI: A. Goobar) starting 58
days after $B-$band maximum.
The SN shows signs of modest host galaxy reddening, with a colour excess of $E(B-V) \approx 0.2$\,mag, 
derived from both optical photometry and high-resolution spectroscopy \citep{silverman2012,munari2013,amanullah2015}.
Using optical light curves, \citet{munari2013} put SN~2012cg at a distance $\mu=30.95$\,mag, close to the
Tully-Fisher estimate in \citet{cortes2008}. By accounting for the reddening and scaling the NIR photometry 
in \citet{amanullah2015} to match SN~2011fe, we adopt a distance modulus of $\mu=30.70 \pm 0.16$\,mag ($D=13.8 \pm 1.0$\,Mpc). 

To compare our limits on CS dust for SN~2014J we also include archival data of the reddened SNe 2006X and 2007le.
{\bf SN~2006X} in M~100 was observed by \spitzer (PI's: P. Meikle and R. Kotak) starting 136 days after $B-$band maximum.
Similar to SN~2014J, 2006X showed signs of non-standard reddening, $E(B - V) \sim 1.4$\,mag with $R_V \sim 1.5$ \citep{wang2008a,folatelli2010}.
{\bf SN~2007le} suffered less extinction than SN~2006X, $E(B-V) \sim 0.39$\,mag, but also had a low $R_V \sim 1.5$ \citep{burns2014}.
The detection of time varying \NaID absorption for both these SNe has been interpreted as being due to CSM at distances $\sim 10^{17}$\,cm 
from the SN \citep{patat2007,simon2009}. Since the SNe are reddened, it has been speculated that dust in the CS environment could play an important role.
We also include data of {\bf SN~2005df} and {\bf SN~2009ig} previously presented in \citet{mcclelland2013} and \citet{gerardy2007}, 
adding two epochs for SN~2009ig at -3 and +36 days from peak brightness, serendipitously observed with \spitzer (PI: K. Sheth).
{\bf SN~2007af} and {\bf SN~2007sr} were observed with \spitzer (PI: R. Kotak) and have well measured optical/near-IR light-curves 
and precise Cepheid distance estimates \citep{riess2011}. 
%%%%%%%%%%%%%%%%%%%%%%%%%%%%%%%%%%%%%%%%%%%%%%%%%%%%%%%%%%%%%%%%%%%%%%%
%%%%%%%%%%%%%%%%%%%%%%%%%%%%%%%%%%%%%%%%%%%%%%%%%%%%%%%%%%%%%%%%%%%%%%%
\section{Mid-IR light curves}
\begin{figure}
\begin{center}
\includegraphics[angle=0,width=1.0\columnwidth]{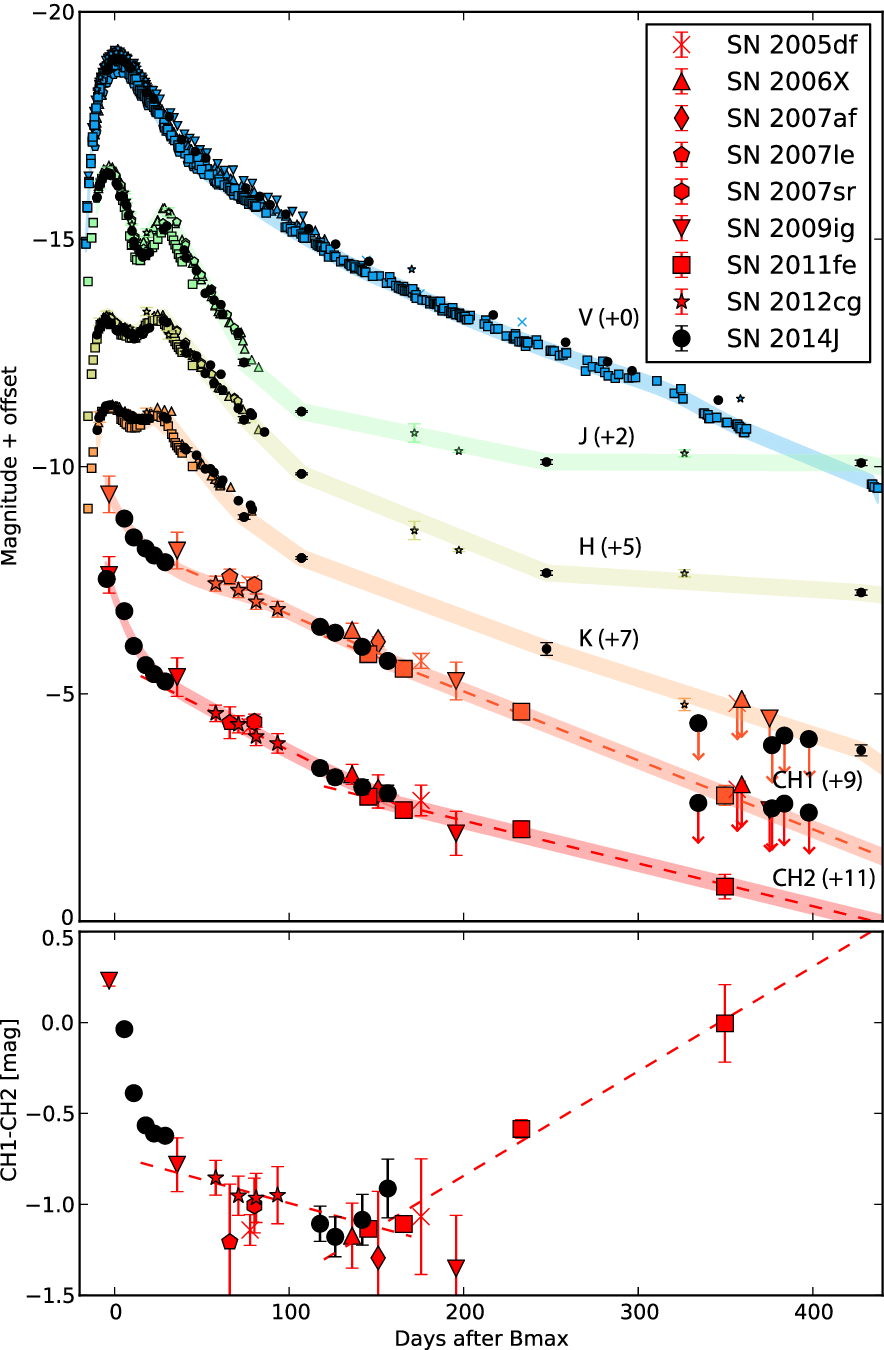}
\caption{Upper panel: Absolute magnitude $V$-band, near-IR ($J, H$ and $K$-band) and mid-IR light-curves 
of the Type Ia SNe used in this study. The magnitudes have been de-reddened using the best fit extinction values 
and shifted by the distance moduli (typically known to $\sim0.2$ mag accuracy) listed in Tab. \ref{tab:sne}. 
The lower panel shows the \spitzer CH1$-$CH2 colour evolution. } \label{fig:lightcurves}
\end{center}
\end{figure}

SN~2014J is the best object to date to build mid-IR light-curve
templates, capturing the full range from before peak brightness to the
nebular phase. To fill in the gaps, caused by limited visibility
windows and scheduling constraints, we make use of archival data of
the SNe~Ia described in Sect. \ref{sec:sne}.  
The composite light-curves shown in Fig. \ref{fig:lightcurves} are
compiled by shifting each SN by their estimated distance modulus and
correcting for host galaxy and Milky Way extinction (see Table~\ref{tab:sne}).
We do not find evidence for variations in the
in the mid-IR light-curve shapes corresponding to the 
different optical decline rates in our sample, ${\Delta}m^{B}_{15}$ = 0.9
to 1.3, suggesting that SNe~Ia are a very homogeneous class
of objects at longer wavelengths. A larger sample of SNe~Ia with multi-epoch mid-IR coverage
is needed to make a more quantitative study.
Three different decline time scales can be recognized in the CH1 and CH2
light-curves.  Although our first detections are before the optical and
near-IR maximum brightness, we can not fully measure the mid-IR light-curve
shapes at these epochs, i.e. fitting the time of maximum in CH1 and
CH2 is impossible. However, fitting the early epochs (-5\,d to 15\,d from $B$-band
maximum) gives linear decline rates of 0.081 and 0.135\,mag\,d$^{-1}$ in CH1 and CH2 respectively, similar
to the decline rates in optical bands.

A break in the mid-IR light-curves occurs $\sim$15 days after $B-$band maximum. 
This roughly coincides with the onset of the secondary maximum in the near-IR bands, 
although data from +15 to +30 days show no signs of a secondary maximum in the mid-IR. 
From 15 days after $B-$band maximum and onwards, the decline rate changes to 1.67 and 1.93 mag/100d 
in CH1 and CH2, respectively.
After $\sim150$ days past $B-$band maximum the linear decline rates of SN~2011fe are
$1.48$\,mag/100d and $0.78$\,mag/100d in CH1 and CH2,
respectively.  The decline rate in CH1 is similar to what is observed
at optical wavelengths, while the decline in CH2 is slower, which can
also be seen in the CH1$-$CH2 color panel in Fig. \ref{fig:lightcurves} as a change towards redder colors.  
\citet{mcclelland2013} analyze the mid-IR
late-time ($>200$ days) decline rates for four SNe (SN~2011fe, SN~2009ig,
SN~2008Q and SN~2005df) and argue that the different decline rates are a
result of doubly ionized elements dominating the bluer CH1 band
(3.6\,$\mu$m) while singly ionized iron-peak
species are controlling the redder CH2 band (4.5\,$\mu$m). 
They also suggest that the interpolated color at +230
days correlates with the light-curve decline rate at maximum
brightness, i.e.\ with $\Delta m^{B}_{15}$. 
%%%%%%%%%%%%%%%%%%%%%%%%%%%%%%%%%%%%%%%%%%%%%%%%%%%%%%%%%%%%%%%%%%%%%%%
%%%%%%%%%%%%%%%%%%%%%%%%%%%%%%%%%%%%%%%%%%%%%%%%%%%%%%%%%%%%%%%%%%%%%%%
\section{Spectral energy distribution}
\begin{figure*}
\begin{center}
   \includegraphics[angle=0,width=1.0\textwidth]{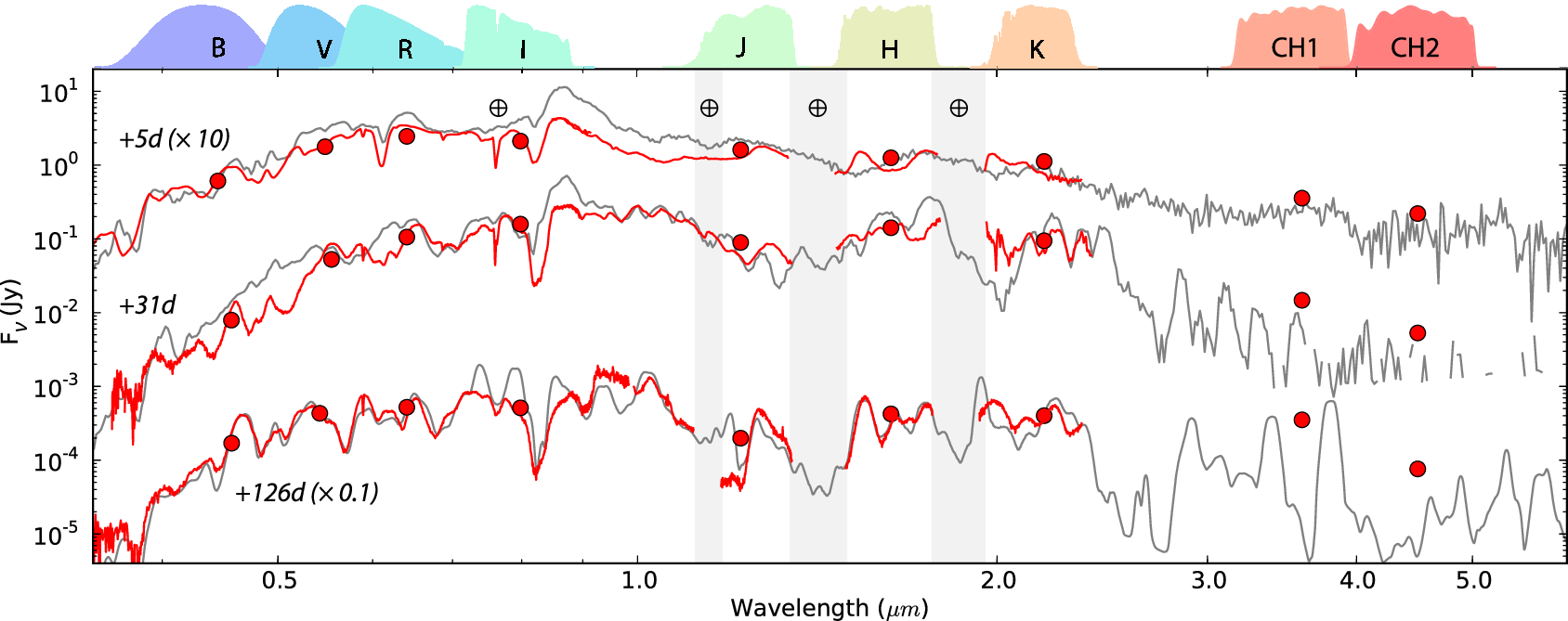}
   \caption{Observed SEDs on SN~2014J at +5, +31 and +126 days past $B$-band
     maximum from optical/NIR spectroscopy (red lines) and broad band $BVRIJHK$ and \spitzer
     CH1 and CH2 photometry (red circles). For comparison, synthetic spectra of the W7 model (gray lines) at similar epochs
    are shown. Vertical, gray bars indicate regions of low atmospheric transmission.}
  \label{fig:sed}
\end{center}
\end{figure*}

Lacking spectra in the wavelength range probed by our \spitzer observations, we examine how the shape of the SED, 
as measured through optical/near-IR spectroscopy together with the \spitzer broad-band observations, matches model spectra from numerical simulations.
In Fig.~\ref{fig:sed} we show observed optical and near-IR spectra (red lines) together with optical, near-IR and mid-IR broad-band 
photometry (red circles) of SN~2014J at three epochs (5, 31 and 126 days after $B$-band maximum).

To this end, we have calculated synthetic spectra for the hydrodynamic
explosion model W7 \citep{nomoto1984}. W7 is known to reproduce the
observed characteristics of normal SNe~Ia at optical
\citep[e.g.][]{branch1985a} and near-IR wavelengths
\citep[e.g.][]{gall2012a}. To obtain synthetic spectra at 5 and 31
days past $B$-band maximum we performed radiative transfer simulations
using the {\sc artis} code \citep{kromer2009a} and the atomic data set
described in \citet{gall2012a}. For the latest epoch (+126 days past
$B$-band maximum) we used a nebular code \citep[described in][]{jerkstrand2011,jerkstrand2014}. 
The flux in the model spectra between $B-$band maximum and +10 days
stems from a mixture of Co\,{\sc ii}, Co\,{\sc iii} and Fe\,{\sc ii}
features and a tiny contribution from intermediate-mass elements.
Starting at 10 days after maximum, the flux in the 2.8 to 3.5 micron region is dominated by singly-ionized iron-group elements.
The lines in the W7 model possibly contributing to the flux in CH1 can
be attributed to Ni\,{\sc ii} (at 2.85, 2.95, 3.11, 3.29 and 3.54\,$\mu$m) and
Fe\,{\sc ii} (3.08\,$\mu$m).

In the late-time model spectrum (+126 days after maximum), the
fluxes in the optical and near-IR 
agree with the observations while the flux levels in the mid-IR are underpredicted.
The dominant lines in CH1 and CH2 are [Fe\,{\sc ii}], with little contribution from [Co\,{\sc ii}].
In CH1 there is also [Fe\,{\sc iii}] (at 2.90 and 3.04\,$\mu$m) and [Co\,{\sc iii}] (at 3.48\,$\mu$m) emission. 

Our observations, shown in Figs.~\ref{fig:lightcurves} and \ref{fig:sed}, bridge the gap
between the late-time near-IR spectra in \citet{friesen2014} and mid-IR spectra of SN~2014J in the
8--13 $\mu$m wavelength range in \citet{telesco2014}. 
\citet{friesen2014} present near-IR spectra of SN~2014J at 70 days past maximum and find that [Ni\,{\sc ii}] fits the emission feature near 1.98\,{$\mu$}m, 
suggesting that a substantial mass of $^{58}$Ni exists near the center of the ejecta, arising from nuclear burning at high density. 
A tentative identification of Mn\,{\sc ii} at 1.15\,{$\mu$}m may support this conclusion as well. 
\citet{telesco2014} compare their observed mid-IR spectra to a delayed
detonation model with $\sim0.6$ M$_{\odot}$ of $^{56}$Ni and claim
that the model is consistent with observations.
Recent multi-dimensional hydrodynamical simulations of Chandrasekhar-mass explosions, however, struggle in producing a 
concentration of stable iron-group elements near the center of the ejecta \citep[e.g.][]{seitenzahl2013a}.
%Reference to SOFIA-paper by \citet{vacca2015} 1-3.4 micron range?}
%%%%%%%%%%%%%%%%%%%%%%%%%%%%%%%%%%%%%%%%%%%%%%%%%%%%%%%%%%%%%%%%%%%%%%%
%%%%%%%%%%%%%%%%%%%%%%%%%%%%%%%%%%%%%%%%%%%%%%%%%%%%%%%%%%%%%%%%%%%%%%%
\section{Constraints on emission from CS dust}
The existence of CS material around individual
nearby SNe~Ia has been suggested by
studies of sodium absorption lines \citep[e.g. SNe 1999cl, 2006X,
2007le, and PTF11kx][]{patat2007,blondin2009,simon2009,dilday2012}.
High-resolution spectra reveal the presence of time-variable and
blueshifted \NaID features, possibly originating from CSM within the
progenitor system.  Studies of large samples of SNe~Ia
\citep{sternberg2011} find that half of all SNe~Ia with detectable \NaID
 absorption at the host-galaxy redshift have \NaID line profiles
with significant blueshifted absorption relative to the strongest
absorption component. This indicates that the absorption occurs in the vicinity of the progenitor systems rather than in the ISM. 
For SN~2014J, high-resolution spectra reveal no signs of time-variable \NaID absorption \citep{goobar2014a,foley2014,welty2014,ritchey2015,maeda2016}, 
consistent with a clean CS environment within $\sim 10^{17}$ cm. However, \citet{graham2015} finds time-variable \KI, 
which might be due to ionization of material at even larger distances ($\sim 10^{19}$ cm).
The wavelength dependent extinction towards
SN~2014J has been measured with great accuracy
using a very wide wavelength range, 0.2 -- 2.2\,$\mu$m,
through a combination of {\it Hubble Space Telescope} and ground-based observations. 
\citet{amanullah2014} found that the reddening can be described with either 
a MW-type extinction law with a low
value of the total-to-selective extinction, $R_{V} = 1.4 \pm 0.1$, i.e.,
corresponding to non-standard dust grains in the ISM, or by
invoking the effective extinction law from \citet{goobar2008}. The latter
arises from multiple scattering of photons on ``normal'' dust grains, but
surrounding the supernova, i.e., in the CS medium, as also
discussed by \citet{wang2005}.   

Non-standard reddening has been noted in studies of individual and
large samples of SNe~Ia. For example, the extinction of SN~2006X was
studied in \citet{folatelli2010}, showing that the reddening is
incompatible with the average extinction law of the Milky Way. Their
findings augmented the large body of evidence indicating that the
reddening of many SNe~Ia show a steeper wavelength dependence ($R_{V}
< 3.1$) than that which is typically observed for stars in our Galaxy.
Previously, \citet{nobili2008} derived $R_{V}=1.75 \pm 0.27$ from a
statistical study of 80 low-redshift SNe~Ia. Similarly, when the
colour-brightness relation is fitted jointly with cosmological
parameters in the SN~Ia Hubble diagram, using a wide range of SN~Ia
redshifts, low values of $R_{V}$ are obtained \citep[see e.g.][for a
recent compilation]{betoule2014}.

\citet{amanullah2011} simulated the impact of thin CS dust shells
located at radii $r_{\rm d} \sim 10^{16}$ -- $10^{19}$\,cm ($\sim
0.003 - 3\,\mathrm{pc}$) from the SN and found that this scenario
would also perturb the optical light-curve shapes and introduce a time
dependent color excess, $\Delta E(B-V) \sim
0.05-0.1$\,mag. \citet{foley2014} claim to have detected a time
variable color excess for SN\,2014J, which led them to conclude
that dimming by CS dust accounts for about half of the
extinction. However, this interpretation has been challenged by \citet{brown2014}. 
By exploring the mid-IR wavelength range, we have a unique way to test if
dust in the CSM plays a significant role in explaining the
non-standard reddening towards SN\,2014J and other highly reddened
normal SNe~Ia.
%%%%%%%%%%%%%%%%%%%%%%%%%%%%%%%%%%%%%%%%%%%%%%%%%%%%%%%%%%%%%%%%%%%%%%%
%%%%%%%%%%%%%%%%%%%%%%%%%%%%%%%%%%%%%%%%%%%%%%%%%%%%%%%%%%%%%%%%%%%%%%%
\subsection{Dust models}
If pre-existing CS dust is the source of non-standard reddening, it will be radiatively
heated by absorption of UV/optical photons from the SN or
collisionally heated by the SN shock. Thermal emission at IR wavelengths could therefore be the ``smoking gun'' for
detecting or ruling out the presence of CS dust.
To model the emission from pre-existing CS dust we
consider the idealized case \citep[described
in][]{hildebrand1983,dwek1985,fox2010} of an optically thin (at mid-IR wavelengths) dust cloud of mass
$M_{\rm d}$ with dust particles of radius $a$, emitting thermally at a
single equilibrium temperature $T_{\rm d}$. The expected flux at a
distance $D$ is
\begin{equation}
  F_{\nu} = M_{d} \frac{\kappa_{\nu}(a) B_{\nu}(T_{\rm d})}{D^{2}} ,
\end{equation}
where $B_{\nu}(T_{\rm d})$ is the Planck blackbody function and
$\kappa_{\nu}(a)$ is the dust mass absorption coefficient.

Since we do not know the nature of the SN~Ia
progenitor systems and their potential dust production mechanisms, we
will consider separate scenarios of either silicate or graphite grains
of radius $a=0.1\,\mu$m and a mixture of silicate and graphitic grains of different sizes 
(MW3.1) that reproduce the standard $R_V=3.1$ Milky Way dust properties \citep[described
in][]{drainelee1984,laordraine1993,weingartnerdraine2001}.

In what follows, we assume that the low-reddening SNe 2011fe and 2012cg represent the intrinsic IR flux from a SN~Ia (dashed lines in Fig.~\ref{fig:lightcurves}), 
to put limits on any additional excess emission from CS dust around the reddened SNe 2014J, 2006X and 2007le.
Taking both instrumental noise and distance modulus uncertainties into account, the differences between the reddened and un-reddened SNe are not statistically significant. 
Hence, we compute $3\sigma$ upper limits on $F_{\rm dust} = F_{\rm red~SN} - F_{\rm unred.~SN}$ from the {\it Spitzer} data, and use Eq.~1 to constrain the dust 
temperature and mass for SN\,2014J ($4.5\mu$m limits shown as red contours in Fig.~\ref{fig:contour_14J}) as well as for other SNe  ($4.5\mu$m limits shown as red 
contours in Fig.~\ref{fig:contour_gra_sne}). The $3.6\,\mu$m limits are comparable, but slightly less constraining than those obtained from the $4.5\,\mu$ data. 
We complement this with a similar analysis using $K-$band data of SNe 2014J and 2006X (green contours in Fig.~\ref{fig:contour_gra_sne}). 

\subsection{Expected emission from heated dust in the thin shell approximation}
\begin{figure}
 \includegraphics[width=80mm]{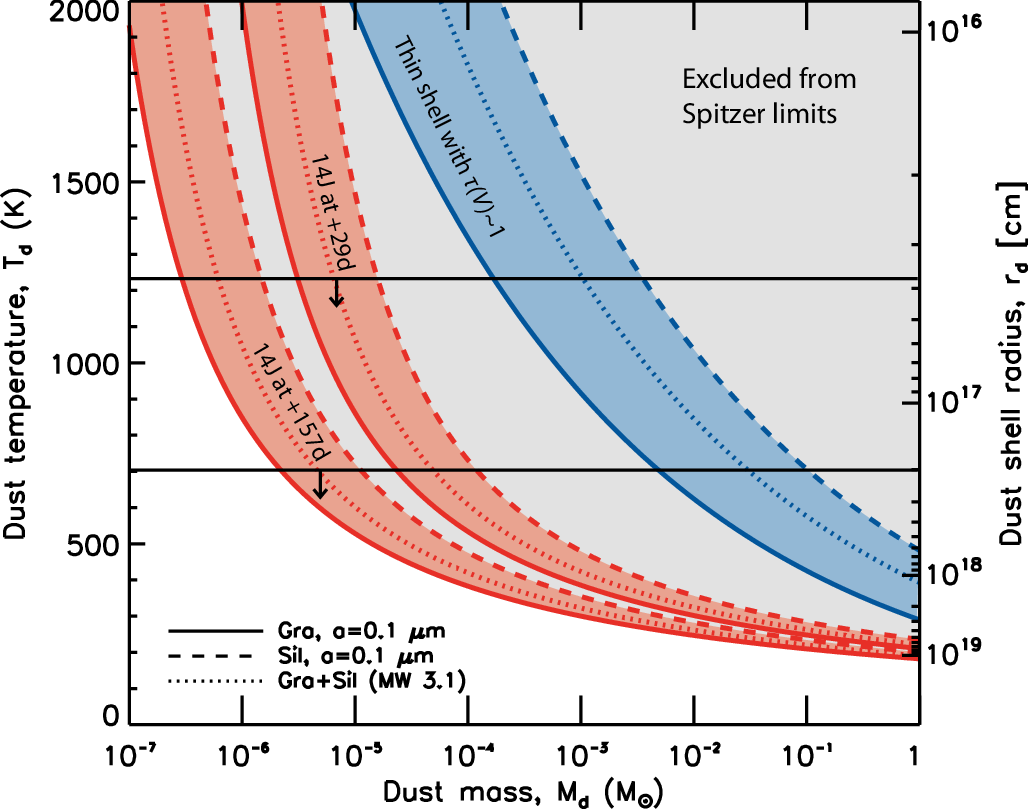}
  \caption{The red lines show 3$\sigma$ upper limits on dust emission around SN~2014J using \spitzer $4.5\,\mu$m data at 
  29 and 157 days after maximum brightness. The solid, dashed and dotted contours indicate limits using the graphitic, 
  silicate and MW3.1 dust models, respectively. Assuming the thin shell approximation, we can estimate the expected dust 
  temperatures and shell radii probed at these epochs (black lines) and the expected dust mass from Eq.~\ref{mexp} (blue lines).}
  \label{fig:contour_14J}
\end{figure}

\begin{figure}
 \includegraphics[width=80mm]{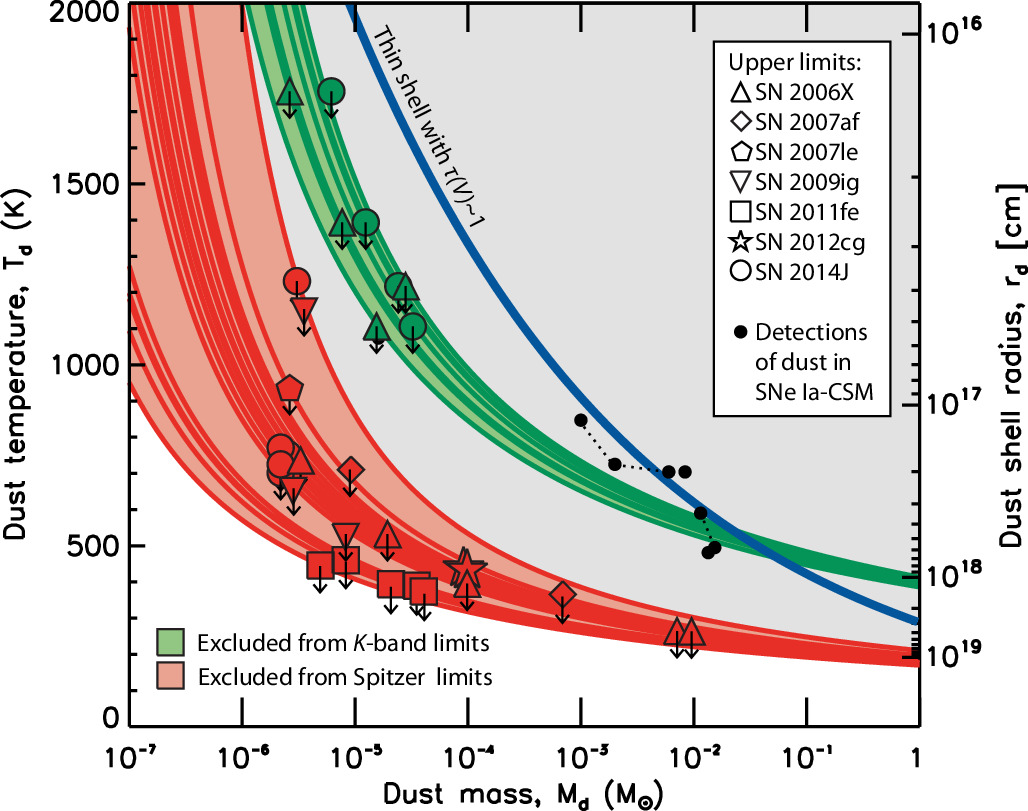}
   \caption{ 3$\sigma$ upper limits on CS dust emission around the reddened SNe 2014J, 2006X and 2007le at different epochs from 
   $K-$band data (green contours and symbols) and \spitzer $4.5\,\mu$m data (red contours and symbols), assuming graphitic dust gains of radius 0.1 $\mu$m. 
   Also shown are the \spitzer limits for SNe 2007af and 2009ig along with the late time non-detections of SNe 2011fe and 2012cg.
  The blue solid line indicates the expected dust mass from Eq.~\ref{mexp}. Black dots are detections of warm dust around Ia-CSM SNe.}
  \label{fig:contour_gra_sne}
\end{figure}

In order to break the degeneracy in the $M_{\rm d}$--$T_{\rm d}$ plane, we make some further assumptions on the dust model.
Although the geometric distribution of CS dust could be complex, we adopt the simple thin shell 
approximation where the dust is distributed uniformly within the shell, to provide an estimate of the expected mid-IR emission from heated dust \citep[comparable 
to the models for reddening proposed in][]{goobar2008,amanullah2011}. This allows us to estimate the expected temperature as a function of shell radius, shown 
as the right-hand side vertical axis of Figs.~\ref{fig:contour_14J} and \ref{fig:contour_gra_sne}.
We estimate the minimal dust shell radius, $r_{\rm d} = ct/2$, that could give rise to a detectable
IR echo at each observed epoch, $t$. Dust at this radius will be heated to
\begin{equation}
  T_{\rm d, exp} \sim 4.0 \left( \frac{L}{a}\right)^{\frac{1}{6}} r_{\rm d}^{-\frac{1}{3}} \,,
\end{equation}  
where we assume that the peak SN bolometric luminosity of $\sim 3 \times 10^{9} \,\mathrm{L}_{\sun}$ is 
heating a spherical dust shell with grain sizes of $a=0.1\,\mu$m \citep{kruegel2003}.
The upper bound on the dust temperature is set by the evaporation
temperature of the dust grains ($T \lsim 2000\, \mathrm{K}$),
corresponding to a minimal dust survival radius $r_{\rm evap} \sim
10^{16}\, \mathrm{cm}$.

Furthermore, in order for a thin dust shell at $r_{\rm d}$ to have significant opacity in
the optical $V-$band, $\tau(V) \sim 1$, the required dust mass can be
estimated from the absorption cross-sections, $\sigma_{\rm abs}(\lambda=0.55
\mu{\rm m})/ m_{\rm dust} = \kappa_{V}$, as
\begin{equation}\label{mexp}
  M_{\rm d,exp} \sim 4\pi r_{\rm d}^{2} \frac{\tau(V)}{\kappa(V)} \,,
\end{equation} 
where $\kappa(V) \sim 5 \cdot 10^{4} {\rm cm}^2 {\,\rm g}^{-1}$ for
graphitic grains, $\kappa(V) \sim 2 \cdot 10^{3} {\rm cm}^2{\,\rm g}^{-1}$
for silicate grains of size $a=0.1\,\mu$m and $\kappa(V) \sim 2 \cdot
10^{3} {\rm cm}^2{\,\rm g}^{-1}$ for the MW3.1 mixture.  Thus, for a thin
spherical dust shell at $r_{\rm d}$ the total dust mass, $M_{\rm d, exp} \sim
10^{-4}$ -- $10^{-5} \left( r_{\rm d} / 10^{16} {\rm cm} \right)^2 {\rm M}_{\odot}$,
depending on dust grain composition, is needed to explain the observed
reddening if it mainly arises in the CS environment. This corresponds to the blue lines in Figs.~\ref{fig:contour_14J} and \ref{fig:contour_gra_sne}.
The black horizontal lines in Fig. 3 indicate the expected dust temperature (and corresponding shell radius) at two of the epochs 
(29 and 157 days from $B$-band maximum) from where the limits are derived for SN\,2014J. Similarly, the symbols in Fig.~\ref{fig:contour_gra_sne} 
indicate the limits derived at different epochs for an extended sample of SNe observed \spitzer. 

Using $K$-band data for SNe 2014J and 2006X, to explore possible emission from hot CS dust close to the exploding star
($T_{\rm d} \gsim 1200$ K for $r_{\rm d} < 5\cdot 10^{16}$ cm), we are able to rule out $M_{d} > 10^{-5}$ M$_{\odot}$.  
These limits are comparable to the results in \citet{maeda2014}. 
By adding the \spitzer data we can significantly reduce the allowed parameter space, excluding
$M_{\rm d} > 10^{-6}$ M$_{\odot}$ ($T_{\rm d} > 700$ K, for
$r_{\rm d} < 2 \cdot 10^{17}$ cm). 

Our combined limits for SNe 2014J correspond to an upper limit on the possible amount of pre-existing dust surrounding the 
SN progenitor, $M_{\rm d} \lsim 10^{-5} \,\mathrm{M}_{\odot}$ within $r_{\rm d} \sim 2 \cdot 10^{17}$ cm, depending on the 
assumed grain size and composition. However, regardless of the specific dust grain composition, the derived upper limits on 
the CS dust mass are significantly lower than what would be required to explain the observed reddening of SN\,2014J, 
in apparent contraction to the claims in \citet{foley2014}. Similar limits are obtained for the reddened SNe 2006X and 2007le (see Fig.~\ref{fig:contour_gra_sne}).
For reference, we also show detections of dust around a subset of peculiar SNe~Ia
interacting strongly with a dense CSM (SNe~Ia-CSM) in
Fig.~\ref{fig:contour_gra_sne}.  \spitzer observations of SNe~Ia-CSM
2002ic, 2005gj and 2008J
\citep{2011ApJ...741....7F,fox2013,taddia2012}
show evidence for late-time ($> 500$ days after maximum brightness) mid-IR emission from
warm ($T_{\rm d} \sim 500$--$800$\,K) dust.  Assuming a simple dust
population of a single size ($a=0.1\,\mu$m) graphitic grains yield dust
masses of 0.006 -- 0.022\,M$_{\sun}$. The dust parameters are most consistent with a
pre-existing dust shell that lies beyond the forward-shock radius,
most likely radiatively heated by optical and X-ray emission
continuously generated by late-time CSM interaction.
%%%%%%%%%%%%%%%%%%%%%%%%%%%%%%%%%%%%%%%%%%%%%%%%%%%%%%%%%%%%%%%%%%%%%%%
%%%%%%%%%%%%%%%%%%%%%%%%%%%%%%%%%%%%%%%%%%%%%%%%%%%%%%%%%%%%%%%%%%%%%%%
\section{Summary and conclusions}
We have analyzed the mid-IR light curves of SN~2014J and several other
SNe~Ia observed with \spitzer, spanning the range from before peak
brightness well into the nebular phase. We have characterized, for the
first time, the SN~Ia light-curve evolution at $3.6$ and $4.5$\,$\mu$m. Our
observations indicate that SNe~Ia form a very homogeneous class of
objects at these wavelengths, possibly without any light-curve shape
variations. In particular, the mid-IR light curves do not show any
evidence for a secondary maximum, as opposed to the case in the
near-IR. The latter was investigated by \citet{kasen2006} to explore
the physics of the exploding system. Extending these studies to the
now probed longer wavelengths should provide critical tests for
SN~Ia models.

The \spitzer observations provide a completely new way to test
models for the circumstellar environment of SNe~Ia and
may help understanding the non-standard reddening measured
both for individual SNe~Ia and in the large SN~Ia samples 
used to derive cosmological parameters.

By comparing the measured mid-IR fluxes at
different epochs for the reddened SNe 2014J, 2006X and 2007le to
unreddened SNe, we can place strong constraints on the emission from
heated dust within $\sim 10^{18}$\,cm from the exploding stars.  
This allows us to exclude the possibility that the bulk of the observed
extinction towards these highly-reddened SNe~Ia is due to CS dust.
\citet{foley2014} claim that half of the extinction ($A_{V} \sim 1$
mag) towards SN~2014J can be attributed to CS dust, while the other
half is due to interstellar dust in M\,82.  However, our limits on dust
emission imply that at most $\tau(V) \approx A_V \lsim 0.1$\,mag of extinction 
can be accounted for by CS dust.
We emphasize that the derived limits are relatively
insensitive to the assumed dust models (as illustrated in
Fig.~\ref{fig:contour_14J}).  Thus, our findings reaffirm the
conclusions from polarization studies of SNe 2014J and 2006X, which
indicate that the dust in the line of sight towards these objects is
most likely of interstellar nature
\citep{kawabata2014,patat2015}. 
Furthermore, the lack of heated material in the CS environment of
SN~2014J is compatible with the non-detection in X-rays and radio
\citep{margutti2014,perez-torres2014}.
The only comparable previous study of searches for emission from heated
circumstellar dust to date was carried out by
\citet{2013MNRAS.431L..43J}, presenting far-IR non-detections of both SNe
2011fe and 2012cg using the {\it Herschel} Photodetector Array Camera and Spectrometer $70\, \mu$m
instrument. These non-detections exclude CS dust masses $M_{\rm d}
\gsim 7 \times 10^{-3}\,\mathrm{M}_{\odot}$ for dust temperatures
$T_{\rm d} \sim 500\,\mathrm{K}$ at a $3\sigma$-level for SN~2011fe,
while the upper limits are one order of magnitude weaker for
SN~2012cg, excluding dust masses $M_{\rm d} \gsim
10^{-1}\,\mathrm{M}_{\odot}$. Thus, our \spitzer study is more than 
two orders of magnitude more sensitive than previous attempts. 

The mid-IR non-detections for SNe 2006X, 2007le, 2011fe and
2012cg at late epochs ($>600$ days after max) can constrain the possible
presence of dust at very large radii ($>10^{18}$\,cm). These limits could be of value to 
models proposed by \citet{soker2014}, which seek to explain the time variable \NaID 
absorption seen in SNe 2006X and 2007le, provide limits on the allowed distance to the 
reflecting dust responsible for light echoes \citep{crotts2014,wang2008b,drozdov2014} 
or constrain the possibility of newly formed dust grains in SN~Ia ejecta \citep{nozawa2011}. 

To summarize, this work expands on previous efforts to study the 
wide wavelength range of the SED of SNe~Ia and provides the first statistical
sample of SNe~Ia in the mid-IR and a detailed study of the highly reddened 
object SN~2014J and its CS environment. 
The non-detection of thermal emission from heated
dust in the CS environment of SN~2014J, 
as well as for SNe 2006X and 2007le
where the detection of circumstellar matter at $10^{17}$\,cm has been 
claimed, argues against the proposed explanation for
the non-standard reddening of SNe~Ia invoking multiple-scattering 
on CS dust \citep{wang2005,goobar2008}. 
Thus, the non-standard reddening may be unrelated to the SN site and originate from the host galaxy ISM
being different than what has been derived for the Milky Way, as indicated by \citet{phillips2013}.
This could have a serious impact for
our understanding of the properties of dust grains in distant galaxies, with profound
implications for essentially all areas of extragalactic astronomy.
%%%%%%%%%%%%%%%%%%%%%%%%%%%%%%%%%%%%%%%%%%%%%%%%%%%%%%%%%%%%%%%%%%%%%%%
%%%%%%%%%%%%%%%%%%%%%%%%%%%%%%%%%%%%%%%%%%%%%%%%%%%%%%%%%%%%%%%%%%%%%%%
\clearpage{}
\section*{Acknowledgements}
We are greatful to Claes Fransson for stimulating discussions. 
RA and AG acknowledge support from the Swedish Research
Council and the Swedish Space Board.  MMK acknowledges
support from the Hubble Fellowship and Carnegie-Princeton Fellowship.
We acknowledge the John von Neumann Institute for Computing %(J\"{u}lich, Germany) 
for granting computing time on the supercomputer JUQUEEN at the J\"{u}lich Supercomputing Centre.
Observations made with the \spitzer telescope, the Nordic 
Optical Telescope, operated by the Nordic Optical Telescope Scientific Association at the
Observatorio del Roque de los Muchachos, La Palma, Spain, of the
Instituto de Astrofisica de Canarias and the Mount Abu 1.2m Infrared 
telescope, India. 
%%%%%%%%%%%%%%%%%%%%%%%%%%%%%%%%%%%%%%%%%%%%%%%%%%%%%%%%%%%%%%%%%%%%%%%
%%%%%%%%%%%%%%%%%%%%%%%%%%%%%%%%%%%%%%%%%%%%%%%%%%%%%%%%%%%%%%%%%%%%%%%
\bibliography{14Jmir_v2}

\begin{thebibliography}{77}
\expandafter\ifx\csname natexlab\endcsname\relax\def\natexlab#1{#1}\fi

\bibitem[{{Amanullah} \& {Goobar}(2011)}]{amanullah2011}
{Amanullah} R., {Goobar} A., 2011, \apj, 735, 20

\bibitem[{{Amanullah} {et~al}\mbox{.}(2014){Amanullah}, {Goobar}, {Johansson},
  {Banerjee}, {Venkataraman}, {Joshi}, {Ashok}, {Cao}, {Kasliwal}, {Kulkarni},
  {Nugent}, {Petrushevska}, \& {Stanishev}}]{amanullah2014}
{Amanullah} R. {et~al.}, 2014, \apjl, 788, L21

\bibitem[{{Amanullah} {et~al}\mbox{.}(2015){Amanullah}, {Johansson}, {Goobar},
  {Ferretti}, {Papadogiannakis}, {Petrushevska}, {Brown}, {Cao}, {Contreras},
  {Dahle}, {Elias-Rosa}, {Fynbo}, {Gorosabel}, {Guaita}, {Hangard}, {Howell},
  {Hsiao}, {Kankare}, {Kasliwal}, {Leloudas}, {Lundqvist}, {Mattila}, {Nugent},
  {Phillips}, {Sandberg}, {Stanishev}, {Sullivan}, {Taddia}, {{\"O}stlin},
  {Asadi}, {Herrero-Illana}, {Jensen}, {Karhunen}, {Lazarevic}, {Varenius},
  {Santos}, {Sridhar}, {Wallstr{\"o}m}, \& {Wiegert}}]{amanullah2015}
{Amanullah} R. {et~al.}, 2015, \mnras, 453, 3300

\bibitem[{{Ashall} {et~al}\mbox{.}(2014){Ashall}, {Mazzali}, {Bersier},
  {Hachinger}, {Phillips}, {Percival}, {James}, \& {Maguire}}]{ashall2014}
{Ashall} C., {Mazzali} P., {Bersier} D., {Hachinger} S., {Phillips} M.,
  {Percival} S., {James} P., {Maguire} K., 2014, \mnras, 445, 4427

\bibitem[{{Betoule} {et~al}\mbox{.}(2014){Betoule}, {Kessler}, {Guy}, {Mosher},
  {Hardin}, {Biswas}, {Astier}, {El-Hage}, {Konig}, {Kuhlmann}, {Marriner},
  {Pain}, {Regnault}, {Balland}, {Bassett}, {Brown}, {Campbell}, {Carlberg},
  {Cellier-Holzem}, {Cinabro}, {Conley}, {D'Andrea}, {DePoy}, {Doi}, {Ellis},
  {Fabbro}, {Filippenko}, {Foley}, {Frieman}, {Fouchez}, {Galbany}, {Goobar},
  {Gupta}, {Hill}, {Hlozek}, {Hogan}, {Hook}, {Howell}, {Jha}, {Le Guillou},
  {Leloudas}, {Lidman}, {Marshall}, {M{\"o}ller}, {Mour{\~a}o}, {Neveu},
  {Nichol}, {Olmstead}, {Palanque-Delabrouille}, {Perlmutter}, {Prieto},
  {Pritchet}, {Richmond}, {Riess}, {Ruhlmann-Kleider}, {Sako}, {Schahmaneche},
  {Schneider}, {Smith}, {Sollerman}, {Sullivan}, {Walton}, \&
  {Wheeler}}]{betoule2014}
{Betoule} M. {et~al.}, 2014, \aap, 568, A22

\bibitem[{{Blondin} {et~al}\mbox{.}(2009){Blondin}, {Prieto}, {Patat},
  {Challis}, {Hicken}, {Kirshner}, {Matheson}, \& {Modjaz}}]{blondin2009}
{Blondin} S., {Prieto} J.~L., {Patat} F., {Challis} P., {Hicken} M., {Kirshner}
  R.~P., {Matheson} T., {Modjaz} M., 2009, \apj, 693, 207

\bibitem[{{Branch} {et~al}\mbox{.}(1985){Branch}, {Doggett}, {Nomoto}, \&
  {Thielemann}}]{branch1985a}
{Branch} D., {Doggett} J.~B., {Nomoto} K., {Thielemann} F.-K., 1985, \apj, 294,
  619

\bibitem[{{Brown} {et~al}\mbox{.}(2015){Brown}, {Smitka}, {Wang}, {Breeveld},
  {de Pasquale}, {Hartmann}, {Krisciunas}, {Kuin}, {Milne}, {Page}, \&
  {Siegel}}]{brown2014}
{Brown} P.~J. {et~al.}, 2015, \apj, 805, 74

\bibitem[{{Burns} {et~al}\mbox{.}(2014){Burns}, {Stritzinger}, {Phillips},
  {Hsiao}, {Contreras}, {Persson}, {Folatelli}, {Boldt}, {Campillay},
  {Castell{\'o}n}, {Freedman}, {Madore}, {Morrell}, {Salgado}, \&
  {Suntzeff}}]{burns2014}
{Burns} C.~R. {et~al.}, 2014, \apj, 789, 32

\bibitem[{{Cort{\'e}s} {et~al}\mbox{.}(2008){Cort{\'e}s}, {Kenney}, \&
  {Hardy}}]{cortes2008}
{Cort{\'e}s} J.~R., {Kenney} J.~D.~P., {Hardy} E., 2008, \apj, 683, 78

\bibitem[{{Crotts}(2015)}]{crotts2014}
{Crotts} A.~P.~S., 2015, \apjl, 804, L37

\bibitem[{{Diehl} {et~al}\mbox{.}(2014){Diehl}, {Siegert}, {Hillebrandt},
  {Grebenev}, {Greiner}, {Krause}, {Kromer}, {Maeda}, {R{\"o}pke}, \&
  {Taubenberger}}]{diehl2014}
{Diehl} R. {et~al.}, 2014, Science, 345, 1162

\bibitem[{{Dilday} {et~al}\mbox{.}(2012){Dilday}, {Howell}, {Cenko},
  {Silverman}, {Nugent}, {Sullivan}, {Ben-Ami}, {Bildsten}, {Bolte}, {Endl},
  {Filippenko}, {Gnat}, {Horesh}, {Hsiao}, {Kasliwal}, {Kirkman}, {Maguire},
  {Marcy}, {Moore}, {Pan}, {Parrent}, {Podsiadlowski}, {Quimby}, {Sternberg},
  {Suzuki}, {Tytler}, {Xu}, {Bloom}, {Gal-Yam}, {Hook}, {Kulkarni}, {Law},
  {Ofek}, {Polishook}, \& {Poznanski}}]{dilday2012}
{Dilday} B. {et~al.}, 2012, Science, 337, 942

\bibitem[{{Draine} \& {Lee}(1984)}]{drainelee1984}
{Draine} B.~T., {Lee} H.~M., 1984, \apj, 285, 89

\bibitem[{{Drozdov} {et~al}\mbox{.}(2015){Drozdov}, {Leising}, {Milne},
  {Pearcy}, {Riess}, {Macri}, {Bryngelson}, \& {Garnavich}}]{drozdov2014}
{Drozdov} D., {Leising} M.~D., {Milne} P.~A., {Pearcy} J., {Riess} A.~G.,
  {Macri} L.~M., {Bryngelson} G.~L., {Garnavich} P.~M., 2015, \apj, 805, 71

\bibitem[{{Dwek}(1985)}]{dwek1985}
{Dwek} E., 1985, \apj, 297, 719

\bibitem[{{Folatelli} {et~al}\mbox{.}(2010){Folatelli}, {Phillips}, {Burns},
  {Contreras}, {Hamuy}, {Freedman}, {Persson}, {Stritzinger}, {Suntzeff},
  {Krisciunas}, {Boldt}, {Gonz{\'a}lez}, {Krzeminski}, {Morrell}, {Roth},
  {Salgado}, {Madore}, {Murphy}, {Wyatt}, {Li}, {Filippenko}, \&
  {Miller}}]{folatelli2010}
{Folatelli} G. {et~al.}, 2010, \aj, 139, 120

\bibitem[{{Foley} {et~al}\mbox{.}(2012){Foley}, {Challis}, {Filippenko},
  {Ganeshalingam}, {Landsman}, {Li}, {Marion}, {Silverman}, {Beaton},
  {Bennert}, {Cenko}, {Childress}, {Guhathakurta}, {Jiang}, {Kalirai},
  {Kirshner}, {Stockton}, {Tollerud}, {Vink{\'o}}, {Wheeler}, \&
  {Woo}}]{foley2012a}
{Foley} R.~J. {et~al.}, 2012, \apj, 744, 38

\bibitem[{{Foley} {et~al}\mbox{.}(2014){Foley}, {Fox}, {McCully}, {Phillips},
  {Sand}, {Zheng}, {Challis}, {Filippenko}, {Folatelli}, {Hillebrandt},
  {Hsiao}, {Jha}, {Kirshner}, {Kromer}, {Marion}, {Nelson}, {Pakmor},
  {Pignata}, {R{\"o}pke}, {Seitenzahl}, {Silverman}, {Skrutskie}, \&
  {Stritzinger}}]{foley2014}
{Foley} R.~J. {et~al.}, 2014, \mnras, 443, 2887

\bibitem[{{Fox} {et~al}\mbox{.}(2010){Fox}, {Chevalier}, {Dwek}, {Skrutskie},
  {Sugerman}, \& {Leisenring}}]{fox2010}
{Fox} O.~D., {Chevalier} R.~A., {Dwek} E., {Skrutskie} M.~F., {Sugerman}
  B.~E.~K., {Leisenring} J.~M., 2010, \apj, 725, 1768

\bibitem[{{Fox} {et~al}\mbox{.}(2011){Fox}, {Chevalier}, {Skrutskie},
  {Soderberg}, {Filippenko}, {Ganeshalingam}, {Silverman}, {Smith}, \&
  {Steele}}]{2011ApJ...741....7F}
{Fox} O.~D. {et~al.}, 2011, \apj, 741, 7

\bibitem[{{Fox} \& {Filippenko}(2013)}]{fox2013}
{Fox} O.~D., {Filippenko} A.~V., 2013, \apjl, 772, L6

\bibitem[{{Friesen} {et~al}\mbox{.}(2014){Friesen}, {Baron}, {Wisniewski},
  {Parrent}, {Thomas}, {Miller}, \& {Marion}}]{friesen2014}
{Friesen} B., {Baron} E., {Wisniewski} J.~P., {Parrent} J.~T., {Thomas} R.~C.,
  {Miller} T.~R., {Marion} G.~H., 2014, \apj, 792, 120

\bibitem[{{Gall} {et~al}\mbox{.}(2012){Gall}, {Taubenberger}, {Kromer}, {Sim},
  {Benetti}, {Blanc}, {Elias-Rosa}, \& {Hillebrandt}}]{gall2012a}
{Gall} E.~E.~E., {Taubenberger} S., {Kromer} M., {Sim} S.~A., {Benetti} S.,
  {Blanc} G., {Elias-Rosa} N., {Hillebrandt} W., 2012, \mnras, 427, 994

\bibitem[{{Gerardy} {et~al}\mbox{.}(2007){Gerardy}, {Meikle}, {Kotak},
  {H{\"o}flich}, {Farrah}, {Filippenko}, {Foley}, {Lundqvist}, {Mattila},
  {Pozzo}, {Sollerman}, {Van Dyk}, \& {Wheeler}}]{gerardy2007}
{Gerardy} C.~L. {et~al.}, 2007, \apj, 661, 995

\bibitem[{{Goobar}(2008)}]{goobar2008}
{Goobar} A., 2008, \apjl, 686, L103

\bibitem[{{Goobar} {et~al}\mbox{.}(2014){Goobar}, {Johansson}, {Amanullah},
  {Cao}, {Perley}, {Kasliwal}, {Ferretti}, {Nugent}, {Harris}, {Gal-Yam},
  {Ofek}, {Tendulkar}, {Dennefeld}, {Valenti}, {Arcavi}, {Banerjee},
  {Venkataraman}, {Joshi}, {Ashok}, {Cenko}, {Diaz}, {Fremling}, {Horesh},
  {Howell}, {Kulkarni}, {Papadogiannakis}, {Petrushevska}, {Sand}, {Sollerman},
  {Stanishev}, {Bloom}, {Surace}, {Dupuy}, \& {Liu}}]{goobar2014a}
{Goobar} A. {et~al.}, 2014, \apjl, 784, L12

\bibitem[{{Goobar} {et~al}\mbox{.}(2015){Goobar}, {Kromer}, {Siverd},
  {Stassun}, {Pepper}, {Amanullah}, {Kasliwal}, {Sollerman}, \&
  {Taddia}}]{goobar2014b}
{Goobar} A. {et~al.}, 2015, \apj, 799, 106

\bibitem[{{Goobar} \& {Leibundgut}(2011)}]{goobar2011}
{Goobar} A., {Leibundgut} B., 2011, Annual Review of Nuclear and Particle
  Science, 61, 251

\bibitem[{{Graham} {et~al}\mbox{.}(2015){Graham}, {Valenti}, {Fulton}, {Weiss},
  {Shen}, {Kelly}, {Zheng}, {Filippenko}, {Marcy}, {Howell}, {Burt}, \&
  {Rivera}}]{graham2015}
{Graham} M.~L. {et~al.}, 2015, \apj, 801, 136

\bibitem[{{Hicken} {et~al}\mbox{.}(2009){Hicken}, {Challis}, {Jha}, {Kirshner},
  {Matheson}, {Modjaz}, {Rest}, {Wood-Vasey}, {Bakos}, {Barton}, {Berlind},
  {Bragg}, {Brice{\~n}o}, {Brown}, {Caldwell}, {Calkins}, {Cho}, {Ciupik},
  {Contreras}, {Dendy}, {Dosaj}, {Durham}, {Eriksen}, {Esquerdo}, {Everett},
  {Falco}, {Fernandez}, {Gaba}, {Garnavich}, {Graves}, {Green}, {Groner},
  {Hergenrother}, {Holman}, {Hradecky}, {Huchra}, {Hutchison}, {Jerius},
  {Jordan}, {Kilgard}, {Krauss}, {Luhman}, {Macri}, {Marrone}, {McDowell},
  {McIntosh}, {McNamara}, {Megeath}, {Mochejska}, {Munoz}, {Muzerolle},
  {Naranjo}, {Narayan}, {Pahre}, {Peters}, {Peterson}, {Rines}, {Ripman},
  {Roussanova}, {Schild}, {Sicilia-Aguilar}, {Sokoloski}, {Smalley}, {Smith},
  {Spahr}, {Stanek}, {Barmby}, {Blondin}, {Stubbs}, {Szentgyorgyi}, {Torres},
  {Vaz}, {Vikhlinin}, {Wang}, {Westover}, {Woods}, \& {Zhao}}]{hicken2009}
{Hicken} M. {et~al.}, 2009, \apj, 700, 331

\bibitem[{{Hildebrand}(1983)}]{hildebrand1983}
{Hildebrand} R.~H., 1983, QJRAS, 24, 267

\bibitem[{{Jerkstrand} {et~al}\mbox{.}(2015){Jerkstrand}, {Ergon}, {Smartt},
  {Fransson}, {Sollerman}, {Taubenberger}, {Bersten}, \&
  {Spyromilio}}]{jerkstrand2014}
{Jerkstrand} A., {Ergon} M., {Smartt} S.~J., {Fransson} C., {Sollerman} J.,
  {Taubenberger} S., {Bersten} M., {Spyromilio} J., 2015, \aap, 573, A12

\bibitem[{{Jerkstrand} {et~al}\mbox{.}(2011){Jerkstrand}, {Fransson}, \&
  {Kozma}}]{jerkstrand2011}
{Jerkstrand} A., {Fransson} C., {Kozma} C., 2011, \aap, 530, A45

\bibitem[{{Johansson} {et~al}\mbox{.}(2013){Johansson}, {Amanullah}, \&
  {Goobar}}]{2013MNRAS.431L..43J}
{Johansson} J., {Amanullah} R., {Goobar} A., 2013, \mnras, 431, L43

\bibitem[{{Kasen}(2006)}]{kasen2006}
{Kasen} D., 2006, \apj, 649, 939

\bibitem[{{Kasliwal} {et~al}\mbox{.}(2017){Kasliwal}, {Bally}, {Masci}, {Cody},
  {Bond}, {Jencson}, {Tinyanont}, {Cao}, {Contreras}, {Dykhoff}, {Amodeo},
  {Armus}, {Boyer}, {Cantiello}, {Carlon}, {Cass}, {Cook}, {Corgan}, {Faella},
  {Fox}, {Green}, {Gehrz}, {Helou}, {Hsiao}, {Johansson}, {Khan}, {Lau},
  {Langer}, {Levesque}, {Milne}, {Mohamed}, {Morrell}, {Monson}, {Moore},
  {Ofek}, {O' Sullivan}, {Parthasarthy}, {Perez}, {Perley}, {Phillips},
  {Prince}, {Shenoy}, {Smith}, {Surace}, {Van Dyk}, {Whitelock}, \&
  {Williams}}]{Kasliwal2017}
{Kasliwal} M.~M. {et~al.}, 2017, ArXiv e-prints

\bibitem[{{Kawabata} {et~al}\mbox{.}(2014){Kawabata}, {Akitaya}, {Yamanaka},
  {Itoh}, {Maeda}, {Moritani}, {Ui}, {Kawabata}, {Mori}, {Nogami}, {Nomoto},
  {Suzuki}, {Takaki}, {Tanaka}, {Ueno}, {Chiyonobu}, {Harao}, {Matsui},
  {Miyamoto}, {Nagae}, {Nakashima}, {Nakaya}, {Ohashi}, {Ohsugi}, {Komatsu},
  {Sakimoto}, {Sasada}, {Sato}, {Tanaka}, {Urano}, {Yamashita}, {Yoshida},
  {Arai}, {Ebisuda}, {Fukazawa}, {Fukui}, {Hashimoto}, {Honda}, {Izumiura},
  {Kanda}, {Kawaguchi}, {Kawai}, {Kuroda}, {Masumoto}, {Matsumoto}, {Nakaoka},
  {Takata}, {Uemura}, \& {Yanagisawa}}]{kawabata2014}
{Kawabata} K.~S. {et~al.}, 2014, \apjl, 795, L4

\bibitem[{{Kromer} \& {Sim}(2009)}]{kromer2009a}
{Kromer} M., {Sim} S.~A., 2009, \mnras, 398, 1809

\bibitem[{{Kruegel}(2003)}]{kruegel2003}
{Kruegel} E., 2003, {The physics of interstellar dust}. Taylor \& Francis

\bibitem[{{Laor} \& {Draine}(1993)}]{laordraine1993}
{Laor} A., {Draine} B.~T., 1993, \apj, 402, 441

\bibitem[{{Maeda} {et~al}\mbox{.}(2015){Maeda}, {Nozawa}, {Nagao}, \&
  {Motohara}}]{maeda2014}
{Maeda} K., {Nozawa} T., {Nagao} T., {Motohara} K., 2015, \mnras, 452, 3281

\bibitem[{{Maeda} {et~al}\mbox{.}(2016){Maeda}, {Tajitsu}, {Kawabata}, {Foley},
  {Honda}, {Moritani}, {Tanaka}, {Hashimoto}, {Ishigaki}, {Simon}, {Phillips},
  {Yamanaka}, {Nogami}, {Arai}, {Aoki}, {Nomoto}, {Milisavljevic}, {Mazzali},
  {Soderberg}, {Schramm}, {Sato}, {Harakawa}, {Morrell}, \&
  {Arimoto}}]{maeda2016}
{Maeda} K. {et~al.}, 2016, \apj, 816, 57

\bibitem[{{Mager} {et~al}\mbox{.}(2013){Mager}, {Madore}, \&
  {Freedman}}]{2013ApJ...777...79M}
{Mager} V.~A., {Madore} B.~F., {Freedman} W.~L., 2013, \apj, 777, 79

\bibitem[{{Margutti} {et~al}\mbox{.}(2014){Margutti}, {Parrent}, {Kamble},
  {Soderberg}, {Foley}, {Milisavljevic}, {Drout}, \& {Kirshner}}]{margutti2014}
{Margutti} R., {Parrent} J., {Kamble} A., {Soderberg} A.~M., {Foley} R.~J.,
  {Milisavljevic} D., {Drout} M.~R., {Kirshner} R., 2014, \apj, 790, 52

\bibitem[{{Marion} {et~al}\mbox{.}(2015){Marion}, {Sand}, {Hsiao}, {Banerjee},
  {Valenti}, {Stritzinger}, {Vink{\'o}}, {Joshi}, {Venkataraman}, {Ashok},
  {Amanullah}, {Binzel}, {Bochanski}, {Bryngelson}, {Burns}, {Drozdov},
  {Fieber-Beyer}, {Graham}, {Howell}, {Johansson}, {Kirshner}, {Milne},
  {Parrent}, {Silverman}, {Vervack}, \& {Wheeler}}]{marion2014}
{Marion} G.~H. {et~al.}, 2015, \apj, 798, 39

\bibitem[{{Marion} {et~al}\mbox{.}(2013){Marion}, {Vinko}, {Wheeler}, {Foley},
  {Hsiao}, {Brown}, {Challis}, {Filippenko}, {Garnavich}, {Kirshner},
  {Landsman}, {Parrent}, {Pritchard}, {Roming}, {Silverman}, \&
  {Wang}}]{marion2013}
{Marion} G.~H. {et~al.}, 2013, \apj, 777, 40

\bibitem[{{Matheson} {et~al}\mbox{.}(2012){Matheson}, {Joyce}, {Allen}, {Saha},
  {Silva}, {Wood-Vasey}, {Adams}, {Anderson}, {Beck}, {Bentz}, {Bershady},
  {Binkert}, {Butler}, {Camarata}, {Eigenbrot}, {Everett}, {Gallagher},
  {Garnavich}, {Glikman}, {Harbeck}, {Hargis}, {Herbst}, {Horch}, {Howell},
  {Jha}, {Kaczmarek}, {Knezek}, {Manne-Nicholas}, {Mathieu}, {Meixner},
  {Milliman}, {Power}, {Rajagopal}, {Reetz}, {Rhode}, {Schechtman-Rook},
  {Schwamb}, {Schweiker}, {Simmons}, {Simon}, {Summers}, {Young}, {Weyant},
  {Wilcots}, {Will}, \& {Williams}}]{matheson2012}
{Matheson} T. {et~al.}, 2012, \apj, 754, 19

\bibitem[{{Mazzali} {et~al}\mbox{.}(2014){Mazzali}, {Sullivan}, {Hachinger},
  {Ellis}, {Nugent}, {Howell}, {Gal-Yam}, {Maguire}, {Cooke}, {Thomas},
  {Nomoto}, \& {Walker}}]{mazzali2014}
{Mazzali} P.~A. {et~al.}, 2014, \mnras, 439, 1959

\bibitem[{{McClelland} {et~al}\mbox{.}(2013){McClelland}, {Garnavich}, {Milne},
  {Shappee}, \& {Pogge}}]{mcclelland2013}
{McClelland} C.~M., {Garnavich} P.~M., {Milne} P.~A., {Shappee} B.~J., {Pogge}
  R.~W., 2013, \apj, 767, 119

\bibitem[{{Milne} {et~al}\mbox{.}(2010){Milne}, {Brown}, {Roming}, {Holland},
  {Immler}, {Filippenko}, {Ganeshalingam}, {Li}, {Stritzinger}, {Phillips},
  {Hicken}, {Kirshner}, {Challis}, {Mazzali}, {Schmidt}, {Bufano}, {Gehrels},
  \& {Vanden Berk}}]{milne2010}
{Milne} P.~A. {et~al.}, 2010, \apj, 721, 1627

\bibitem[{{Munari} {et~al}\mbox{.}(2013){Munari}, {Henden}, {Belligoli},
  {Castellani}, {Cherini}, {Righetti}, \& {Vagnozzi}}]{munari2013}
{Munari} U., {Henden} A., {Belligoli} R., {Castellani} F., {Cherini} G.,
  {Righetti} G.~L., {Vagnozzi} A., 2013, New Astronomy, 20, 30

\bibitem[{{Nobili} \& {Goobar}(2008)}]{nobili2008}
{Nobili} S., {Goobar} A., 2008, \aap, 487, 19

\bibitem[{{Nomoto} {et~al}\mbox{.}(1984){Nomoto}, {Thielemann}, \&
  {Yokoi}}]{nomoto1984}
{Nomoto} K., {Thielemann} F.-K., {Yokoi} K., 1984, \apj, 286, 644

\bibitem[{{Nozawa} {et~al}\mbox{.}(2011){Nozawa}, {Maeda}, {Kozasa}, {Tanaka},
  {Nomoto}, \& {Umeda}}]{nozawa2011}
{Nozawa} T., {Maeda} K., {Kozasa} T., {Tanaka} M., {Nomoto} K., {Umeda} H.,
  2011, \apj, 736, 45

\bibitem[{{Patat} {et~al}\mbox{.}(2007){Patat}, {Chandra}, {Chevalier},
  {Justham}, {Podsiadlowski}, {Wolf}, {Gal-Yam}, {Pasquini}, {Crawford},
  {Mazzali}, {Pauldrach}, {Nomoto}, {Benetti}, {Cappellaro}, {Elias-Rosa},
  {Hillebrandt}, {Leonard}, {Pastorello}, {Renzini}, {Sabbadin}, {Simon}, \&
  {Turatto}}]{patat2007}
{Patat} F. {et~al.}, 2007, Science, 317, 924

\bibitem[{{Patat} {et~al}\mbox{.}(2013){Patat}, {Cordiner}, {Cox}, {Anderson},
  {Harutyunyan}, {Kotak}, {Palaversa}, {Stanishev}, {Tomasella}, {Benetti},
  {Goobar}, {Pastorello}, \& {Sollerman}}]{patat2013}
{Patat} F. {et~al.}, 2013, \aap, 549, A62

\bibitem[{{Patat} {et~al}\mbox{.}(2015){Patat}, {Taubenberger}, {Cox}, {Baade},
  {Clocchiatti}, {H{\"o}flich}, {Maund}, {Reilly}, {Spyromilio}, {Wang},
  {Wheeler}, \& {Zelaya}}]{patat2015}
{Patat} F. {et~al.}, 2015, \aap, 577, A53

\bibitem[{{Pereira} {et~al}\mbox{.}(2013){Pereira}, {Thomas}, {Aldering},
  {Antilogus}, {Baltay}, {Benitez-Herrera}, {Bongard}, {Buton}, {Canto},
  {Cellier-Holzem}, {Chen}, {Childress}, {Chotard}, {Copin}, {Fakhouri},
  {Fink}, {Fouchez}, {Gangler}, {Guy}, {Hillebrandt}, {Hsiao}, {Kerschhaggl},
  {Kowalski}, {Kromer}, {Nordin}, {Nugent}, {Paech}, {Pain}, {P{\'e}contal},
  {Perlmutter}, {Rabinowitz}, {Rigault}, {Runge}, {Saunders}, {Smadja}, {Tao},
  {Taubenberger}, {Tilquin}, \& {Wu}}]{pereira2013}
{Pereira} R. {et~al.}, 2013, \aap, 554, A27

\bibitem[{{P{\'e}rez-Torres} {et~al}\mbox{.}(2014){P{\'e}rez-Torres},
  {Lundqvist}, {Beswick}, {Bj{\"o}rnsson}, {Muxlow}, {Paragi}, {Ryder},
  {Alberdi}, {Fransson}, {Marcaide}, {Mart{\'{\i}}-Vidal}, {Ros}, {Argo}, \&
  {Guirado}}]{perez-torres2014}
{P{\'e}rez-Torres} M.~A. {et~al.}, 2014, \apj, 792, 38

\bibitem[{Phillips {et~al}\mbox{.}(2013)Phillips, Simon, Morrell, Burns, Cox,
  Foley, Karakas, Patat, Sternberg, Williams, Gal-Yam, Hsiao, Leonard, Persson,
  Stritzinger, Thompson, Campillay, Contreras, Folatelli, Freedman, Hamuy,
  Roth, Shields, Suntzeff, Chomiuk, Ivans, Madore, Penprase, Perley, Pignata,
  Preston, \& Soderberg}]{phillips2013}
Phillips M.~M. {et~al.}, 2013, The Astrophysical Journal, 779, 38

\bibitem[{{Riess} {et~al}\mbox{.}(2011){Riess}, {Macri}, {Casertano},
  {Lampeitl}, {Ferguson}, {Filippenko}, {Jha}, {Li}, \& {Chornock}}]{riess2011}
{Riess} A.~G. {et~al.}, 2011, \apj, 730, 119

\bibitem[{{Ritchey} {et~al}\mbox{.}(2015){Ritchey}, {Welty}, {Dahlstrom}, \&
  {York}}]{ritchey2015}
{Ritchey} A.~M., {Welty} D.~E., {Dahlstrom} J.~A., {York} D.~G., 2015, \apj,
  799, 197

\bibitem[{{Schweizer} {et~al}\mbox{.}(2008){Schweizer}, {Burns}, {Madore},
  {Mager}, {Phillips}, {Freedman}, {Boldt}, {Contreras}, {Folatelli},
  {Gonz{\'a}lez}, {Hamuy}, {Krzeminski}, {Morrell}, {Persson}, {Roth}, \&
  {Stritzinger}}]{schweizer2008}
{Schweizer} F. {et~al.}, 2008, \aj, 136, 1482

\bibitem[{{Seitenzahl} {et~al}\mbox{.}(2013){Seitenzahl},
  {Ciaraldi-Schoolmann}, {R{\"o}pke}, {Fink}, {Hillebrandt}, {Kromer},
  {Pakmor}, {Ruiter}, {Sim}, \& {Taubenberger}}]{seitenzahl2013a}
{Seitenzahl} I.~R. {et~al.}, 2013, \mnras, 429, 1156

\bibitem[{{Silverman} {et~al}\mbox{.}(2012){Silverman}, {Ganeshalingam},
  {Cenko}, {Filippenko}, {Li}, {Barth}, {Carson}, {Childress}, {Clubb},
  {Cucchiara}, {Graham}, {Marion}, {Nguyen}, {Pei}, {Tucker}, {Vinko},
  {Wheeler}, \& {Worseck}}]{silverman2012}
{Silverman} J.~M. {et~al.}, 2012, \apjl, 756, L7

\bibitem[{{Simon} {et~al}\mbox{.}(2009){Simon}, {Gal-Yam}, {Gnat}, {Quimby},
  {Ganeshalingam}, {Silverman}, {Blondin}, {Li}, {Filippenko}, {Wheeler},
  {Kirshner}, {Patat}, {Nugent}, {Foley}, {Vogt}, {Butler}, {Peek},
  {Rosolowsky}, {Herczeg}, {Sauer}, \& {Mazzali}}]{simon2009}
{Simon} J.~D. {et~al.}, 2009, \apj, 702, 1157

\bibitem[{{Simon} {et~al}\mbox{.}(2007){Simon}, {Gal-Yam}, {Penprase}, {Li},
  {Quimby}, {Silverman}, {Allende Prieto}, {Wheeler}, {Filippenko}, {Martinez},
  {Beeler}, \& {Patat}}]{simon2007}
{Simon} J.~D. {et~al.}, 2007, \apjl, 671, L25

\bibitem[{{Soker}(2014)}]{soker2014}
{Soker} N., 2014, \mnras, 444, L73

\bibitem[{{Sternberg} {et~al}\mbox{.}(2011){Sternberg}, {Gal-Yam}, {Simon},
  {Leonard}, {Quimby}, {Phillips}, {Morrell}, {Thompson}, {Ivans}, {Marshall},
  \& {Filippenko}}]{sternberg2011}
{Sternberg} A. {et~al.}, 2011, Science, 333, 856

\bibitem[{{Taddia} {et~al}\mbox{.}(2012){Taddia}, {Stritzinger}, {Phillips},
  {Burns}, {Heinrich-Josties}, {Morrell}, {Sollerman}, {Valenti}, {Anderson},
  {Boldt}, {Campillay}, {Castellon}, {Contreras}, {Folatelli}, {Freedman},
  {Hamuy}, {Krzeminski}, {Leloudas}, {Maeda}, {Persson}, {Roth}, \&
  {Suntzeff}}]{taddia2012}
{Taddia} F. {et~al.}, 2012, \aap, 545, L7

\bibitem[{{Telesco} {et~al}\mbox{.}(2015){Telesco}, {H{\"o}flich}, {Li},
  {{\'A}lvarez}, {Wright}, {Barnes}, {Fern{\'a}ndez}, {Hough}, {Levenson},
  {Mari{\~n}as}, {Packham}, {Pantin}, {Rebolo}, {Roche}, \&
  {Zhang}}]{telesco2014}
{Telesco} C.~M. {et~al.}, 2015, \apj, 798, 93

\bibitem[{{Wang}(2005)}]{wang2005}
{Wang} L., 2005, \apjl, 635, L33

\bibitem[{{Wang} {et~al}\mbox{.}(2008{\natexlab{a}}){Wang}, {Li}, {Filippenko},
  {Foley}, {Smith}, \& {Wang}}]{wang2008b}
{Wang} X., {Li} W., {Filippenko} A.~V., {Foley} R.~J., {Smith} N., {Wang} L.,
  2008{\natexlab{a}}, \apj, 677, 1060

\bibitem[{{Wang} {et~al}\mbox{.}(2008{\natexlab{b}}){Wang}, {Li}, {Filippenko},
  {Krisciunas}, {Suntzeff}, {Li}, {Zhang}, {Deng}, {Foley}, {Ganeshalingam},
  {Li}, {Lou}, {Qiu}, {Shang}, {Silverman}, {Zhang}, \& {Zhang}}]{wang2008a}
{Wang} X. {et~al.}, 2008{\natexlab{b}}, \apj, 675, 626

\bibitem[{{Weingartner} \& {Draine}(2001)}]{weingartnerdraine2001}
{Weingartner} J.~C., {Draine} B.~T., 2001, \apj, 548, 296

\bibitem[{{Welty} {et~al}\mbox{.}(2014){Welty}, {Ritchey}, {Dahlstrom}, \&
  {York}}]{welty2014}
{Welty} D.~E., {Ritchey} A.~M., {Dahlstrom} J.~A., {York} D.~G., 2014, \apj,
  792, 106

\end{thebibliography}
%%%%%%%%%%%%%%%%%%%%%%%%%%%%%%%%%%%%%%%%%%%%%%%%%%%%%%%%%%%%%%%%%%%%%%%
%%%%%%%%%%%%%%%%%%%%%%%%%%%%%%%%%%%%%%%%%%%%%%%%%%%%%%%%%%%%%%%%%%%%%%%
\clearpage
\begin{table*}
\begin{minipage}{126mm}
\caption{Type Ia supernovae observed in the mid-IR}\label{tab:sne}
  \begin{tabular}{@{}llllcccclcl@{}}
    \hline
Target	&  $t_{B,{\rm max}}$	& Host		&  Distance		&Distance		& $E(B-V)$ &$R_V$ & $\Delta m^{B}_{15}$   & Reference \\
               	&  MJD			& galaxy		&  modulus (mag)	&(Mpc)	    	& (mag)	  &		  & (mag)			& 		\\
    \hline\hline
2005df 	& 53598.4 	& NGC\,1559 	& 31.26 (0.14)		& 15.7 (2.3) 	& 0.03 	  	& 3.1		& 1.20 			& 1, 2, 3 		\\
2006X 	& 53785.6 	& NGC\,4321 	& 30.91 (0.14) 		& 15.2 (1.0) 	& 1.24	  	& 1.6 	& 1.29			& 4, 5 		\\
2007af	& 54173.8		& NGC\,5584	& 31.72 (0.07)		& 22.0 (0.7)	& 0.04	  	& 3.0		& 1.20			& 6, 7, 8 		\\
2007le 	& 54398.9	 	& NGC\,7721	& 32.24 (0.16) 		& 28.1 (2.1) 	& 0.39	  	& 1.5  	& 1.10			& 9 			\\
2007sr	& 54448.3		& NGC\,4038	& 31.66 (0.08)		& 21.5 (0.8)	& 0.13	  	&1.6		& 0.97			& 6, 10 		\\
2009ig 	& 55080.9 	& NGC\,1015 	& 32.82 (0.09)		& 36.6 (1.5) 	& $<$0.05  	& -		& 0.89			& 2, 11, 12  	\\
2011fe 	& 55815.8  	& NGC\,5457 	& 28.93 (0.16)		& 6.1 (0.5) 	& 0.03	  	& 3.1		& 1.11			& 2, 13, 14 	\\
2012cg 	& 56082.0		& NGC\,4424 	& 30.70 (0.16)		& 13.8 (1.0) 	& 0.15		& 2.6	 	& 0.89			& 14, 15,16 	\\
2014J	& 56689.2 	& M\,82		& 27.60 (0.16)  		& 3.3 (0.15)	& 1.37   	  	& 1.4 	& 0.95			& 17, 18, 19 	\\
   \hline\hline
     \end{tabular}\\
 	(1):~\citet{gerardy2007}, (2): ~\citet{mcclelland2013}, (3):~\citet{milne2010}, (4):~\citet{wang2008a}, (5):~\citet{folatelli2010}
	(6):~\citet{riess2011}, (7):~\citet{simon2007}, (8):~\citet{hicken2009}, (9):~\citet{burns2014}, (10):~\citet{schweizer2008}, (11):~\citet{foley2012a}, (12):~\citet{marion2013}, (13):~\citet{matheson2012}, (14):~\citet{munari2013}, (15):~\citet{silverman2012}, (16):~\citet{amanullah2015}, (17):~\citet{amanullah2014}, (18):~\citet{foley2014}, (19):~\citet{marion2014}, (20):~\citet{amanullah2015}
\end{minipage}
\end{table*}

\begin{table*}
\begin{minipage}{126mm}
\caption{Log of spectroscopic observations of SN~2014J}\label{tab:obslog}
  \begin{tabular}{@{}lllcccclcl@{}}
    \hline
UT date			&  MJD		& Days from			&  Observatory /	& Wavelength range	&  Exp. time 		& Airmass	 	\\ 
               			& 			& $B_{\rm max}$		&  Instrument		& ($\mu$m)		& (s)	    			& 		  	\\
    \hline\hline	
2014-02-06$^a$ 	& 56694.9 	& +5.1				& NOT/ALFOSC	& 0.35 -- 0.90		& 180		& 1.61		\\
2014-02-06$^a$	& 56695.0		& +5.2				& Mt Abu			& 0.85 -- 2.35		& 1200		& 1.59		\\
2014-03-03		& 56718.9		& +29.1				& Mt Abu			& 0.85 -- 2.35		& 1080		& 1.48		\\
2014-03-05 		& 56721.1 	& +31.3				& NOT/ALFOSC	& 0.35 -- 0.90		& 180		& 1.35		\\	
2014-06-07		& 56815.3		& +125.5				& Keck/MOSFIRE	& 1.15 -- 2.35		& 480		& 2.19		\\
2014-06-09 		& 56817.2 	& +127.4				& ARC/DIS		& 0.35 -- 0.95		& 120		& 1.73		\\
   \hline\hline	
     \end{tabular}\\
 	$^a$ Published in \citet{marion2014}.
	\end{minipage}
\end{table*}
%%%%%%%%%%%%%%%%%%%%%%%%%%%%%%%%%%%%%%%%%%%%%%%%%%%%%%%%%%%%%%%%%%%%%%%
%%%%%%%%%%%%%%%%%%%%%%%%%%%%%%%%%%%%%%%%%%%%%%%%%%%%%%%%%%%%%%%%%%%%%%%
\clearpage
\begin{table*}
\begin{centering}
\begin{minipage}{126mm}
  \caption{Spitzer observations and photometry of Type Ia supernovae in the mid-IR. }\label{tab:photometry}
  \begin{tabular}{@{}llrccccccl@{}}
    \hline
SN		&  MJD			& Epoch			& $F_{\nu}^{\rm CH1}$ 	& $F_{\nu}^{\rm CH2}$	& CH1	& CH2	&  Sub?	\\
               	& 				& (days)			& ($\mu$Jy)			& ($\mu$Jy)			& (mag)	& (mag)	 &	(y/n)		\\
    \hline\hline
2005df 	& 53676.0 	&   77.6 	&  327.0 (6.7) 	& 73.2 (5.7) 	& 14.835 (0.022) 	& 15.976 (0.082) 	&  y  \\ 
2005df 	& 53774.0 	&  175.6 	&   68.3 (5.6) 	& 16.4 (5.3) 	& 16.535 (0.085) 	& 17.602 (0.307) 	&  y  \\ 
2005df 	& 53955.1 	&  356.7 	&   13.3 (5.2) 	&  4.6 (5.3)  	& $>$18.136		& $>$17.642 		&  y  \\ 
\hline
2006X 	& 53922.2 	&  136.1 	&157.9 (8.0) 	&   35.3 (6.0) 	& 15.626 (0.054) 	&   16.768 (0.171) 	&  y  \\ 
2006X 	& 54145.4 	&  359.3 	&19.7 (8.0) 	&   13.3 (6.0) 	& $>$17.750	 	&   $>$17.572 		&  y  \\ 
2006X 	& 54662.7 	&  876.6 	& -0.7 (7.6) 	&    1.8 (5.9) 	& $>$17.731	 	&   $>$17.520 		&  y  \\ 
2006X 	& 56722.8 	& 2936.7 	& -6.9 (8.3) 	&   -1.3 (6.5) 	& $>$17.633	 	&   $>$17.415 		&  y  \\ 
2006X 	& 56751.9 	& 2965.8 	& 1.6 ( 8.6) 	&    0.2 (6.3) 	& $>$17.586		&   $>$17.447 		&  y  \\ 
\hline
2007af	& 54324.9		& 151.1	& 62.2 (4.6)   	& 12.8 (5.0)	& 16.594 (0.081)	& 17.865 (0.357)	&  y  \\ 
2007af	& 55280.4		& 1106.6	& 2.4 (5.0)   	& 2.2 (5.0)		& $>$18.181		& $>$17.696 		&  y  \\ 
\hline
2007le	& 54464.7		& 66.0	& 152.0 (5.0)   	& 32.0 (5.0)	& 15.667 (0.059)	& 16.873 (0.312)	&  y  \\ 
\hline
2007sr	& 54528.1		& 80.1	& 220.0 (8.0)   & 55.7 (10.0)	& 15.265 (0.146)	& 16.271 (0.146)	&  y  \\ 
2007sr	& 55616.4		& 1168.4	& 3.3 (8.0)   	&19.2 (8.0)	& $>$17.671		& $>$17.186		&  y  \\ 
\hline
2009ig 	& 55076.9	 	& -3.1 	& 580.7 (10.0) 	& 459.1 (10.0) 	&   14.211 (0.019) 	& 13.982 (0.023) 	&  n  \\ 
2009ig 	& 55115.9	 	& 35.8 	& 186.1 (8.0) 	& 57.9 (8.0) 	&   15.447 (0.046) 	& 16.229 (0.140) 	&  n  \\ 
2009ig 	& 55275.7	 	& 195.6 	& 13.3 (1.4)	& 2.4 (0.7)		&   18.316 (0.110) 	& 19.668 (0.270) 	&  n  \\ 
2009ig 	& 55455.2	 	& 375.2 	& 2.9 (1.1) 	& 1.5 (0.8)		&   $>$19.876 		& $>$19.672		&  n  \\ 
\hline
2011fe 	& 55960.7 	&  145.4 	&  670.6 (8.5) 	&  150.9 (6.4) 	& 14.055 (0.014) 	& 15.190 (0.045) 	 &  y \\
2011fe 	& 55981.0 	&  165.6 	&  498.2 (8.1) 	&  114.9 (6.3) 	& 14.378 (0.018) 	& 15.486 (0.058) 	 &  y \\
2011fe 	& 56048.4 	&  233.0 	&  209.2 (7.0) 	&   78.1 (5.7) 	& 15.320 (0.036) 	& 15.905 (0.076) 	 &  y \\
2011fe 	& 56165.0 	&  349.7 	&   38.3 (5.7) 	&   24.4 (5.4) 	& 17.163 (0.151) 	& 17.168 (0.217) 	 &  y \\
2011fe 	& 56337.1 	&  521.8 	&    3.3 (5.6) 	&   14.1 (6.1) 	& $>$18.058	   	& 17.763 (0.390) 	 &  y \\
2011fe 	& 56348.1 	&  532.8 	&    2.5 (5.8) 	&   15.6 (5.0) 	& $>$18.020	   	& 17.654 (0.302) 	 &  y \\
2011fe 	& 56393.8 	&  578.5 	&    4.4 (5.1) 	&    7.8 (7.7) 	& $>$18.160	   	& $>$17.227	 	&  y \\
2011fe 	& 56452.6 	&  637.3 	&    3.9 (6.0) 	&   -2.6 (4.6) 	& $>$17.983	   	& $>$17.787	 	&  y \\
2011fe 	& 56742.8 	&  927.5 	&   - 			&    2.4 (5.7) 	& - 			  	& $>$17.554	 	&  y \\
2011fe 	& 56771.8 	&  956.5 	&   - 			&    1.9 (5.8) 	& - 		 	   	& $>$17.534	 	&  y \\
2011fe 	& 56902.0 	& 1086.7 	&    3.1 (7.4) 	&   -8.9 (11.2) 	& $>$17.755	   	& $>$16.820		 &  y \\ 
\hline
2012cg	& 56139.9 	& 57.9  	& 549.0 (20.3) 		& 160.0 (13.3)   	& 14.272 (0.039)   	& 15.126 (0.087) 	&  y \\ 
2012cg	& 56152.9   	& 70.9  	& 481.0 (19.0) 		& 128.0 (12.3)   	& 14.416 (0.042)   	& 15.368 (0.100) 	&  y \\ 
2012cg	& 56163.1  	& 81.1  	& 380.0 (20.4) 		& 100.0 (11.9)   	& 14.672 (0.057)   	& 15.636 (0.122) 	&  y \\ 
2012cg	& 56175.4  	& 93.4  	& 326.0 (20.6)  		& 87.0 (12.2)   		& 14.838 (0.067)   	& 15.788 (0.142) 	&  y \\ 
2012cg	& 56723.8    	& 641.8   	& 11.9  (17.2)  		& 6.2 (12.6)   		& $>$16.837		& $>$16.694		&  y \\ 
2012cg	& 56751.9    	& 669.9   	& 4.1 (18.4)   		& 2.2 (12.7)  		& $>$16.764		& $>$16.682 		&  y \\ 
\hline
2014J	& 56685.4  	&  -4.5  	&  - 				& 42273.0 (200.0)	&  -				&  9.071 (0.005)	&  y \\ 
2014J	& 56695.4 	& 5.6 	&  35895.9 (154.9) 	& 22130.1 (97.1)	&   9.734 (0.005)	&  9.774 (0.005) 	&  y \\ 
2014J	& 56700.8 	& 11.0 	&  24508.9 (149.2) 	& 10554.5 (95.9)	& 10.148 (0.007)	&  10.578 (0.010) 	&  y \\ 
2014J	& 56707.6  	& 17.8 	&  18295.8 (117.4) 	& 7130.0 (73.1)		& 10.466 (0.007)  	&  11.004 (0.011) 	&  y \\ 
2014J	& 56712.4 	& 22.6 	&  16491.0 (121.1) 	& 5932.9 (75.2) 	& 10.578 (0.008)	&  11.203 (0.014) 	&  y \\ 
2014J	& 56718.8 	& 29.0 	&  14325.6 (129.4) 	& 5145.3 (80.5) 	& 10.731 (0.010) 	&  11.358 (0.017) 	&  y \\ 
2014J	& 56807.5 	& 117.7 	&  3812.2  (127.3)  	& 876.4 (80.3) 		& 12.168 (0.036)	&  13.280 (0.095) 	&  y \\ 
2014J	& 56816.2 	& 126.4 	&  3273.6  (122.5)  	& 766.0 (75.8) 		& 12.334 (0.040)	&  13.426 (0.102) 	&  y \\ 
2014J	& 56831.7 	& 141.9 	&  2338.2  (121.3)  	& 600.4 (79.9) 		& 12.699 (0.055)	&  13.690 (0.136) 	&  y \\ 
2014J	& 56846.3 	& 156.5 	&  1922.3	 (126.5)  	& 448.6 (79.9) 		& 12.912 (0.069) 	&  14.007 (0.178) 	&  y \\ 
2014J	& 57024.2 	& 334.4 	&  203.3 (130.2)  	& 212.6 (77.6) 		& $>$14.482		&  $>$13.997 		&  y\\ 
2014J	& 57066.4 	& 376.6	&  1.3 (131.8)  		& 165.7 (79.3) 		& $>$14.601		&  $>$14.116 		&  y \\ 
2014J	& 57073.5 	& 383.7 	&  78.7 (132.1)  	& 206.2 (79.4) 		& $>$14.498		&  $>$14.013 		&  y \\ 
2014J	& 57087.5 	& 397.7 	&  48.8 (132.0)  	& 132.3 (79.3) 		& $>$14.694		&  $>$14.209 		&  y \\ 
   \hline\hline
     \end{tabular}
\end{minipage}
\end{centering}
\end{table*}
\label{lastpage}
\end{document}